# Surface currents associated with external kink modes in tokamak plasmas during a major disruption

C. S. Ng[1,2,a)], and A. Bhattacharjee[1]

[1]*Princeton Plasma Physics Laboratory, Princeton University, Princeton, NJ 08543, USA*

[2]*University of Alaska Fairbanks, Fairbanks, Alaska 99775, USA*

Abstract

The surface current on the plasma-vacuum interface during a disruption event involving kink instability can play an important role in driving current into the vacuum vessel. However, there have been disagreements over the nature or even the sign of the surface current in recent theoretical calculations based on idealized step-function background plasma profiles. We revisit such calculations by replacing step-function profiles with more realistic profiles characterized by a strong but finite gradient along the radial direction. It is shown that the resulting surface current is no longer a delta-function current density, but a finite and smooth current density profile with an internal structure, concentrated within the region with a strong plasma pressure gradient. Moreover, this current density profile has peaks of both signs, unlike the delta-function case with a sign opposite to, or the same as the plasma current. We show analytically and numerically that such current density can be separated into two parts, with one of them, called the convective current density, describing the transport of the background plasma density by the displacement, and the other part that remains, called the residual current density. It is argued that consideration of both types of current density is important and can resolve past controversies.

[a]cng2@alaska.edu

I. INTRODUCTION

A major disruption can have serious damaging effects on a tokamak fusion device. It is one of the main design and operational challenges of ITER.[1,2] For a recent discussion of the status of research toward the ITER disruption mitigation system (DMS), see Ref. 3. One important damaging effect is the force on the vacuum vessel due to the large amount of current, commonly called the halo current, flowing into the vessel wall. For a recent discussion on halo current, see Ref. 4. Frequently, a disruption is initiated by a vertical displacement event (VDE). The safety factor at the plasma edge is then reduced with the consequence that an external kink mode can become unstable. Recently, it was proposed that a strong surface current caused by external kink instability, with a sign opposite to that of the main plasma current, can enter the wall and induce a strong force when the plasma touches the wall. Such a current has been called the Hiro current, and has been argued to have an important qualitative effect on the evolution of a disruption.[5-7] The main objective of this paper is to provide a better physical understanding of surface currents, in sign, magnitude, and spatial structure, since there has been significant controversy, without adequate resolution, of the technical issues involved in the analysis, computation, and experimental validation of such currents.[8-11]

One common treatment in these theoretical studies is the use of step-function background current and plasma density profiles, i.e., both quantities drop to zero at a given radial position, the plasma radius. In this paper, we try to concentrate on resolving this issue, which is central to some ongoing controversies on disruption physics, by relaxing the step-function approximation. The outline of this paper is as follows. In Sec. II, we will review the step-function results to describe the problem mathematically in detail, and to introduce notations so that the step-function results can be compared with our new results more easily. In Sec. III, we present some of our calculations based on smooth spatial profiles of current and mass density. In particular, we will show that the integration of the first-order current density of the unstable mode agrees with the surface current calculated in Ref. 9. However, we will also show in Sec. IV that the surface current found in Refs. 5 and 8 corresponds to a part of the first-order current not included in Ref. 9, and termed the residual current in this paper. In Sec. V, we present numerical results showing that these two kinds of current density can be spatially separated, depending on the safety factor, as well as whether the background current and mass density have the same profile. The question of whether the wall surface current has one sign or the other does not have a unique answer, since the answer will depend on potential cancellations of current densities of different signs in the nonlinear regime. Discussion and conclusion are given in Sec. VI.

II. REVIEW OF STEP-FUNCTION RESULTS

In this section, we will review the results using step-function current density and mass density profiles. This serves the purpose of introducing notations, as well as writing down results to be compared with our new calculations with the step-function approximation relaxed. Let us start by following the analytic model in Sec. IV of Ref. 9, using basically the same notations. A cylindrical geometry using coordinates $(r,\theta,\phi)$ is used for simplicity, where $r$ is the distance from the center on a cross-section of constant toroidal angle $\phi$, and $\theta$ is the poloidal angle. Under the reduced MHD representation, in normalized units, the magnetic field is given by

$$\mathbf{B} = \nabla\psi \times \hat{\phi} + B_0\hat{\phi}, \tag{1}$$

where $B_0$ is the uniform toroidal background magnetic field strength, and $\psi$ is the flux function. There is a thin resistive wall at $r = b$, with resistivity $\eta_w$ and thickness $\delta$, which is assumed to be much smaller than $b$, but large enough such that the wall time $\tau_w = \delta b / m \eta_w$ can still be large by choosing a small enough $\eta_w$, with $m$ being a positive integer for the poloidal mode number of an unstable mode. The plasma initially is assumed to have a uniform normalized mass density $\rho = 1$, and a uniform current density $j_{\phi 0} = -\nabla_\perp^2 \psi_0 = -2B_0 / q_0 R$, for $r \leq a < b$, with the safety factor $q_0$ being a constant within this region, and $R$ representing physically the major radius of the device and thus should be much larger than the minor radius $b$ in the cylindrical approximation. Outside the plasma ($r > a$) is assumed to be vacuum with $\rho = 0$ and $j_{\phi 0} = 0$. The background magnetic field of the form of $\mathbf{B}_0 = B_{\theta 0} \hat{\theta} + B_0 \hat{\phi}$ is given generally by

$$B_{\theta 0} = -\frac{\partial \psi_0}{\partial r} = \frac{1}{r} \int_0^r j_{\phi 0}(r') r' dr'. \tag{2}$$

Note that the safety factor $q(r)$, as a function of radial position, is defined by $q(r) \equiv |rB_0 / RB_{\theta 0}|$. More specifically for the step-function $j_{\phi 0}$, we have

$$B_{\theta 0} = \begin{cases} -B_0 r / q_0 R & \text{for } r \leq a \\ -B_0 a^2 / q_0 Rr & \text{for } r > a \end{cases}. \tag{3}$$

Since $B_{\theta 0}$ is continuous over $r = a$, there is no surface current in the original equilibrium (zeroth-order).

We seek a linearly unstable eigenmode with a displacement of the form

$$\boldsymbol{\xi} = \nabla \Phi \times \hat{\phi} \propto e^{i(m\theta + n\phi) + \gamma t}, \tag{4}$$

where $\Phi$ is the stream function, $n$ is the toroidal mode number, and $\gamma > 0$ is the growth rate of the unstable mode. The linear eigenmode satisfies the following equations:

$$\gamma^2 \nabla \cdot (\rho \nabla_\perp \Phi) = \mathbf{B}_0 \cdot \nabla \nabla_\perp^2 \psi_1 + \mathbf{B}_1 \cdot \nabla \nabla_\perp^2 \psi_0, \tag{5}$$

$$\psi_1 = \mathbf{B}_0 \cdot \nabla \Phi = i B_0 k_\parallel \Phi, \tag{6}$$

where

$$k_\parallel = \frac{1}{R}\left[n - \frac{m}{q(r)}\right] = \frac{1}{R}\left[n + \frac{mR}{B_0 r^2}\int_0^r j_{\phi 0} r' dr'\right] = \begin{cases} \dfrac{-1}{q_0 R}[m - n q_0] & \text{for } r < a \\ \dfrac{1}{q_0 R}\left[n q_0 - \dfrac{m a^2}{r^2}\right] & \text{for } r > a \end{cases}, \tag{7}$$

and

$$\mathbf{B}_1 = \nabla \psi_1 \times \hat{\phi} = \frac{im\psi_1}{r}\hat{r} - \frac{\partial \psi_1}{\partial r}\hat{\theta}. \tag{8}$$

Using the step-function current density profile, we have

$$\gamma^2 \nabla \cdot \left[\rho \nabla_\perp \frac{\psi_1}{k_\parallel}\right] = -B_0^2 k_\parallel \nabla_\perp^2 \psi_1 + \frac{2m B_0^2}{q_0 R r}\delta(r - a)\psi_1. \tag{9}$$

Again based on the step-function profile in $\rho$, this equation can be solved by

$$\nabla_\perp^2 \psi_1 = 0 \text{ for } 0 \leq r < a, \ a < r < b, \ b < r. \tag{10}$$

Imposing continuity of $\psi$ at $r = a$ and $r = b$, and boundary conditions of $\psi_1 = 0$ at $r = 0$ and $r \to \infty$, we have

$$\psi_1 = \begin{cases} \psi_p = \psi_{1a}(r/a)^m, & \text{for } 0 \leq r < a \\ \psi_v = \psi_{2a}(r/a)^m + \psi_{3a}(a/r)^m, & \text{for } a < r < b \\ \psi_x = \psi_{4a}(b/r)^m, & \text{for } b < r \end{cases}. \tag{11}$$

Integrating Eq. (9) over $r = a$, we obtain

$$\psi_p = \psi_v, \quad \frac{\gamma^2}{B_0^2}\psi'_p = k_\parallel^2(\psi'_v - \psi'_p) - \frac{2mk_\parallel}{q_0 Ra}\psi_p. \qquad (12)$$

At $r = b$, applying the Ampère Law and Ohm's Law, we have

$$\psi_x = \psi_v, \quad \gamma\delta\psi_x = \eta_w(\psi'_x - \psi'_v). \qquad (13)$$

From these equations, we can obtain an equation for the growth rate $\gamma$,

$$\frac{\gamma^2}{2B_0^2} = -\frac{[1+2/(\gamma\tau_w)]k_\parallel^2}{1-(a/b)^{2m}+2/(\gamma\tau_w)} - \frac{k_\parallel}{Rq_0} = \frac{m-nq_0}{R^2q_0^2}\left\{1 - \frac{(m-nq_0)[1+2/(\gamma\tau_w)]}{1-(a/b)^{2m}+2/(\gamma\tau_w)}\right\}. \qquad (14)$$

This is a third-order algebraic equation for $\gamma$ and so it can be solved analytically. However, it is also illustrative to consider further simplifications in the limit of $\tau_w \to \infty$ for a highly conducting wall:

$$\text{for } nq_0 - m + 1 - (a/b)^{2m} > 0, \quad \frac{\gamma^2}{2B_0^2} \approx \frac{(m-nq_0)[nq_0 - m + 1 - (a/b)^{2m}]}{R^2q_0^2[1-(a/b)^{2m}]}, \qquad (15)$$

which is independent of $\tau_w$, and

$$\text{for } nq_0 - m + 1 - (a/b)^{2m} < 0, \quad \gamma \approx \frac{2(nq_0 - m + 1)}{[m - nq_0 - 1 + (a/b)^{2m}]\tau_w} \to 0, \qquad (16)$$

which decreases inversely with $\tau_w$ and thus is very small. We should point out for completeness that Eqs. (15) and (16) do not apply in the close vicinity of the relation $1-(a/b)^{2m} = m - nq_0$, where $\gamma \approx (2a^m B_0 / b^m Rq_0)^2 / \tau_w^{1/3}$, which lies in between Eqs. (15) and (16) for large $\tau_w$. The analytic solution for $\gamma$ given by Eq. (14) as a function of $q_0$ is plotted on Fig. 4 as the solid curve, for the $m = n = 1$ mode, with $a = 0.5$, $b = 1$, $R = 3$, $B_0 = 1$, $\tau_w = 1000$. This curve shows the behavior as indicated in Eqs. (15) and (16). The condition for instability from Eqs. (14)-(16) is thus

$$m - 1 < nq_0 < m, \tag{17}$$

but in fact for large $\tau_w$, the growth rate is large only for

$$m - 1 + (a/b)^{2m} < nq_0 < m. \tag{18}$$

On the thin conducting wall, there is a current (eddy current) flowing in the toroidal direction with surface current density

$$K_b = B_{1\theta}\big|_{r=b-0^+}^{r=b+0^+} = \psi'_v - \psi'_x = \frac{2(m - nq_0)(a/b)^{m+1} B_0 \xi_r}{q_0 R\left[1 - (a/b)^{2m} + 2/(\gamma\tau_w)\right]}, \tag{19}$$

where the first equality represents a jump condition at $r = b$, $\xi_r = im\Phi/r = im\Phi/a$ is the radial component of the displacement from Eq. (4), $\psi'_v$ is the $r$ derivative of the first-order perturbation of $\psi$ at $r$ slightly less than $b$, and $\psi'_x$ is the $r$ derivative of the first-order perturbation of $\psi$ at $r$ slightly greater than $b$. This current density is always positive, opposite to the negative plasma current, for positive $\xi_r$, i.e., on the side of the plasma moving towards the wall. Although our notation is somewhat different, Eq. (19) is of the same form as Eq. (13) of Ref. 6 in the $\gamma\tau_w \to \infty$ limit, and Eq. (12) of Ref. 7 for the special case of $m = n = 1$. It was pointed out in the above two papers that in the marginally unstable case with $1 \gg m - nq_0 > 0$, this eddy current term can be small due to the multiplicative $m - nq_0$ factor in the numerator of Eq. (19).

Due to the step-function profiles, there is also a surface current density on the plasma surface of $r = a$. It is of the form

$$K = B_{1\theta}\Big|_{r=a-0^+}^{r=a+0^+} = \psi'_p - \psi'_v = -\frac{2B_0}{q_0 R} \frac{(m-nq_0)[1+2/(\gamma\tau_w)]}{[1-(a/b)^{2m}+2/(\gamma\tau_w)]} \xi_r, \quad (20)$$

which is of the same sign as the background plasma current for positive $\xi_r$.

The above analysis up to Eq. (20) is similar to that given in Ref. 9. However, the surface quantity defined by Eq. (20) differs from that used in other papers, e.g., Refs. 7 and 8 which, in our notation, can be written in the form

$$I = K - j_{\phi 0}\xi_r = \frac{2B_0}{q_0 R}\left\{1 - \frac{(m-nq_0)[1+2/(\gamma\tau_w)]}{1-(a/b)^{2m}+2/(\gamma\tau_w)}\right\}\xi_r. \quad (21)$$

By comparing with Eq. (14), we can rewrite $I = q_0 R \gamma^2 \xi_r / B_0(m-nq_0)$, which is small for a small growth rate except when $m-nq_0$ is also small. Equation (21) shows that $I$ is of the sign opposite to that of the background plasma current for positive $\xi_r$, and thus opposite to the sign of $K$ as well.

Since the surface current $I$ is proportional to the plasma displacement $\xi_r$ as shown in Eq. (21), a major criticism of numerical simulations, such as in Ref. 9, is that the use of the boundary condition $\mathbf{V} \cdot \hat{n} = 0$ invalidates the calculation of the surface current.[12] Apparently, this is because when the plasma is in contact with the wall, the restriction $\mathbf{V} \cdot \hat{n} = 0$ will also restrict the correct calculation of the surface current $I$ by Eq. (21). In turn, it is argued that the use of such a boundary condition will preclude the correct determination of the current flowing from the surface of the plasma into the wall (called the Hiro current in Refs. 6 and 7). We will return to address this issue at the end of Sec. IV, after we have presented our understanding of the physical meaning of $K$ and $I$ based on the calculations presented in Sec. III.

III. RESULTS FOR CONTINUOUS PROFILES

We apply the same model as in Sec. II, except that we now use continuous and differentiable profiles for $j_{\phi 0}$, and $\rho$, instead of step-functions. The eigenmode equation is then generalized from Eq. (9) to

$$\gamma^2\left[\rho\nabla_\perp^2\left(\frac{\psi_1}{k_\parallel}\right) + \frac{\partial\rho}{\partial r}\frac{\partial}{\partial r}\left(\frac{\psi_1}{k_\parallel}\right)\right] = -B_0^2 k_\parallel \nabla_\perp^2 \psi_1 + \frac{mB_0}{r}\frac{\partial j_{\phi 0}}{\partial r}\psi_1. \tag{22}$$

The stream function $\Phi$ is still given by Eq. (6) with $k_\parallel$, which is a function of $r$ now, specified by the first two equalities of Eq. (7). While this treatment allows general profiles for $j_{\phi 0}$ and $\rho$ as functions of radial position, we consider in our calculations profiles that are close enough to a step-function in order to make comparison with previous results easier. More specifically, we consider profiles for $j_{\phi 0}$ and $\rho$ that have large gradients at the "boundaries" $a$ and $a_\rho$ with characteristic boundary-layer widths of the order of $1/\kappa$ and $1/\kappa_\rho$, respectively. In general, $a$ and $a_\rho$, $\kappa$, and $\kappa_\rho$ do not have to be the same. These profiles have asymptotic values given by

$$j_{\phi 0} \to \begin{cases} -2B_0/q_0 R & \text{for } \kappa(a-r) \gg 1 \\ 0 & \text{for } \kappa(r-a) \gg 1 \end{cases}, \quad \rho \to \begin{cases} 1 & \text{for } \kappa_\rho(a_\rho - r) \gg 1 \\ 0 & \text{for } \kappa_\rho(r - a_\rho) \gg 1 \end{cases}. \tag{23}$$

To be specific, our calculations are based on $j_{\phi 0}$ of the form

$$j_{\phi 0} = -\frac{B_0}{q_0 R}\text{erfc}[\kappa(r-a)], \tag{24}$$

with $a = 0.5$, $b = 1$. In most calculations we will use the same functional form with $a = a_\rho$ and $\kappa = \kappa_\rho$. Figure 1 shows the plots of normalized profiles given by Eq. (24) for all $\kappa$ values used in our calculations to illustrate how large the spatial gradient is at the "boundary" $a = 0.5$, and that the series of profiles tend to a step-function profile in the limit of large $\kappa$. Using such profiles $k_\parallel$ can be calculated by Eq. (7). Figure 2 shows $k_\parallel$ as functions of $r$ for the case with $m = n = 1$, $R = 3$, $\kappa = 40$, and all values of $q_0$ used in our calculations. We can also see from this figure that $k_\parallel$ is zero somewhere in the range of $b > r > a$ for cases with $1 > q_0 > 0.25$. The point $k_\parallel = 0$ corresponds to the point $q(r) = 1$. Assuming that $\psi_1$ is non-zero at this point, regularity of the stream function $\Phi$ is ensured by requiring that the density $\rho$ vanishes at the same point (where there is no plasma). Physically, an internal kink instability occurs at $k_\parallel = 0$ and $\rho \neq 0$.[13-15] Since we are focusing on effects due to the external kink instability, we ignore the internal kink instability in this paper. Note that this subtlety already exists in the step-function case but is not as apparent since $\rho$ is set exactly zero in the region of $b > r > a$. Figure 3 shows $q_0$ and $\kappa$ parameters for all cases we run in which unstable eigenmodes are found. Each run is indicated by a symbol. Same symbol indicates same $\kappa$. The solid curve is drawn to show that for a $\kappa$ value larger than for the solid curve, the density $\rho$ at the point $k_\parallel = 0$ is less than $0.5\mathrm{erfc}(5) \sim 7.7\mathrm{e}{-13}$. We see that all values of $\kappa$ used in our study satisfy this condition. Other runs with smaller $\kappa$ for the same $q_0$ do not find unstable eigenmodes.

Using the same profiles as in Eq. (24) for both $j_{\phi 0}$, and $\rho$, the growth rates $\gamma$ of unstable eigenmodes can be found. The left panel of Fig. 4 shows $\gamma$ for all cases, with the solid curve indicating $\gamma$ for the step-function case found by solving the third-order algebraic equation as in Eq. (14). The agreement is generally very good. The analytic properties as indicated in Eqs. (15) and (16) are evident. For example, $\gamma$ is small for cases with $q_0 < (a/b)^2 = 0.25$ since now it is inversely proportional to $\tau_w$, which is large ($\tau_w = 1000$). These cases with such small $q_0$ are uncommon in practice and are included here for completeness and confirmation of the analytical results. The right panel of Fig. 4 plots $\gamma$ for the case of $q_0 = 0.7$, for $\kappa = 60$, 100, and 200, showing that in the large $\kappa$ limit, $\gamma$ tends to the step-function ($\kappa \to \infty$) value as indicated by the dotted line, as it should.

Unlike the step-function case, now the first-order current density $j_{\phi 1}$ is not a delta-function (surface current at $r = a$), but distributed smoothly as a function of $r$, concentrated near the boundary layer where the spatial gradient is large. We will show plots of $j_{\phi 1}$ for different cases later, but first we will compare with step-function results by integrating $j_{\phi 1}$ over $r$. With our choice of $j_{\phi 0}$ being negative, we will see that $j_{\phi 1}$ is also mostly negative for positive $\xi_r$, especially for small $q_0$. In Fig. 5, the solid curve shows the analytical values of the surface current $K$ for the step-function case given by Eq. (20) with $\xi_r = -1$, so that $K$ is positive. The data points on this curve are calculated from

$$K = \int_0^b j_{\phi 1} dr = -\int_0^b \nabla_\perp^2 \psi_1 dr \;, \tag{25}$$

for all cases, with the normalization of $\xi_r = -1$ at $r = a$. Therefore, our results confirm that the integrated first-order current density is consistent with the surface current given by Eq. (20) for the step-function case. We also consider the quantity $I = K - j_{\phi 0}\xi_r$ as in Eq. (21), called the surface current by some researchers. The dashed curve in Fig. 5 shows $-I$ (since $I$ is negative for $\xi_r = -1$) by Eq. (21). The data points on it are obtained by subtracting the integrated first-order current $K$ by Eq. (25) from $j_{\phi 0}\xi_r$. The log scale is used for the vertical axis to show good agreement even when $K$ or $-I$ are small. Therefore, we see that the first-order current for continuous cases tend to the surface current $K$ defined by Eq. (20), rather than $I$ defined by Eq. (21), in the step-function limit.

IV. PHYSICAL MEANING OF SURFACE CURRENTS $K$ AND $I$

We have demonstrated by direct calculations that the surface current $K$ defined by Eq. (20) for the step-function case corresponds to the integration of first-order current density $j_{\phi 1}$ through Eq. (25). Mathematically, in the step function limit, we can write $j_{\phi 1} = K\delta(r-a)$. To see the physical meaning of $I$, defined by Eq. (21), we write $j_I = I\delta(r-a)$ so that Eq. (21) becomes

$$j_I = j_{\phi 1} - j_{\phi 0}\xi_r \delta(r-a) = j_{\phi 1} + \xi_r \frac{\partial j_{\phi 0}}{\partial r} \equiv j_{\phi 1} - j_c,  \qquad (26)$$

where we have defined the convective current density through the expression

$$j_c = -\xi_r \frac{\partial j_{\phi 0}}{\partial r},  \qquad (27)$$

and have used the step-function nature of $j_{\phi 0}$ so that its $r$ derivative is proportional to a delta-function. The reason we name $j_c$ as the convective current density is apparent by considering

$$j_{\phi 0}(r) + j_c(r) = j_{\phi 0}(r) - \xi_r \frac{\partial j_{\phi 0}}{\partial r} \approx j_{\phi 0}(r - \xi_r),  \qquad (28)$$

where we have applied a Taylor expansion in the last approximation. This means that the addition of $j_c$, a part of the first-order current density, has the effect of moving $j_{\phi 0}$ in the direction of $\xi_r$. We note that current density in MHD is not a physical quantity convected by the flow, even in the ideal limit when the magnetic field lines are frozen-in. Therefore, the physical meaning of $j_I$ or $I$ can be assigned from Eq. (26) as what remains of the first-order current density after subtracting the convective current density. Due to this physical meaning, we name it the residual current density. While it might seem somewhat arbitrary to separate the first-order current density into two parts, our treatment here in separating into convective current density and residual current density serves the purpose of clearing up confusion in the literature. Furthermore, as we show in Sec. V, for continuous spatial profiles, these two parts of the current density can be spatially separated as distinct peaks of opposite signs.

The analytic form of both $j_c$ and $j_I$ can be obtained from the eigenequation Eq. (22) in the step-function limit by solving the first-order current density and keeping only dominant terms involving $\partial^2 k_\parallel / \partial r^2$ and $\partial \rho / \partial r$. We can then write

$$j_{\phi 1} = -\nabla_\perp^2 \psi_1 \approx j_c + j_I , \tag{29}$$

where it can be shown, with the use of Eq. (7), that the convective current density is given by

$$j_c = -iB_0 \frac{\partial^2 k_\parallel}{\partial r^2} \Phi \approx -\xi_r \frac{\partial j_{\phi 0}}{\partial r} , \tag{30}$$

so that the residual current density is of the form

$$j_I = iB_0 k_\parallel \gamma^2 \frac{\partial \rho}{\partial r} \frac{\partial \Phi}{\partial r} \bigg/ \left( \gamma^2 \rho + B_0^2 k_\parallel^2 \right). \tag{31}$$

A few properties of the residual current density $j_I$ can be observed from these analytical expressions. First, we see that $j_I$ is proportional to $\partial \rho / \partial r$ while $j_c$ is proportional to $\partial^2 k_\parallel / \partial r^2$, and because of the additional dependence on the denominator in Eq. (31), the two do not necessarily have the same radial profile, even for the case in which $j_{\phi 0}$ and $\rho$ have the same profile. We will show profiles of $j_c$ and $j_I$ for some of our cases in Sec. V. By this functional dependence, $j_I$ is therefore exactly zero for the special case with uniform density. In fact, it is interesting to note that for uniform density ($\rho = 1$) and the particular case of $m = n = 1$, which we have been considering so far, Eq. (22) can be solved exactly by

$$\nabla_\perp^2 \Phi = 0, \text{ or } \Phi \propto r, \ \psi_1 = iB_0 k_\parallel \Phi \propto k_\parallel r . \tag{32}$$

This is the case in which the first-order current density is totally given by the convective current density. We have confirmed such a solution expressed in Eq. (32) through our eigenmode calculation. However, note that unstable modes exist only for $q_0 < (a/b)^2 = 0.25$ for this case. Therefore this is an uncommon case in practice and serves only the purpose of a benchmark here, with the possibility of an internal kink instability ignored as mentioned above.

We also see from Eq. (31) that $j_I$ is proportional to $\gamma^2$ and thus is small when the growth rate is small, except in the limit of $q_0 \to m/n$ when $k_\parallel$ is also small by Eq. (7), since then the denominator also becomes small in Eq. (31). This behavior can also be seen by comparing $I$ shown in Fig. 5 and $\gamma$ shown in Fig. 4. Finally, Eq. (31) shows that $j_I$ is proportional to $\partial \Phi / \partial r$. By Eq. (4), this means that $j_I$ is proportional to $\xi_\theta$, rather than $\xi_r$ as in Eq. (21). This has implications for the controversy over the boundary condition $\mathbf{V} \cdot \hat{n} = 0$,[12] which implies that $\xi_r = 0$ at the boundary. Whereas the latter condition would make $I = 0$ by Eq. (21), $j_I$ would not necessarily be zero since $\xi_\theta \neq 0$ on the boundary in general.

V. PROFILES AND RELATIVE IMPORTANCE OF $K$ AND $I$

Using our definitions of convective current density $j_c$ and residual current density $j_I$ through Eqs. (29)-(31), we now show their radial profiles for different cases with continuous background profiles in order to assess their relative importance.

As discussed at the end of Sec. IV, $j_I$ is small for small $\gamma^2$. So from Fig. 4 and Eq. (16) we know that $j_c$ dominates over $j_I$ for $q_0 < (a/b)^2 = 0.25$. This is true for both the uniform and non-uniform density cases. We show the first case with $q_0$ slightly above this range to see if this is still true. Figure 6 shows a case with $q_0 = 0.3$, $a = a_\rho = 0.5$, $b = 1$, $\kappa = \kappa_\rho = 40$, $m = n = 1$, and $R = 3$. The growth rate $\gamma$ for this case is shown in Fig. 4 and is not very small since $\gamma$ increases rapidly for $q_0 > 0.25$. Note that we only plot over a range of $r$ around $r = a$ to show the boundary layer in more detail. The solid curve in the left panel shows $\psi_1$ of the unstable eigenmode. A dotted curve is plotted on the same graph showing the step-function solution solved from Eq. (11). We see that the two solutions agree well except around $r = a$, over the boundary layer. The solid curve in the right panel shows the first-order current density $j_{\phi 1}$. On the same graph, the dashed curve shows the convective current density $j_c$ by Eq. (30), and the dotted curve shows the residual current density $j_I$ by Eq. (31). The addition of these two gives a profile very close to $j_{\phi 1}$. (We do not actually show this in the figure because the sum of the two contributions is virtually indistinguishable from $j_{\phi 1}$.) This case, and other cases not shown here, show that $j_{\phi 1} \approx j_c$ for cases of small $q_0$ i.e., $q_0 < 0.4$.

For larger $q_0$, $j_I$ becomes larger and comparable with $j_c$. Figure 7 shows a case with $q_0 = 0.7$, $\kappa = \kappa_\rho = 100$, and other parameters the same as in Fig. 6. Note that the range of $r$ plotted is different in these two figures. Because of the larger negative values of $j_I$ and the shift of its profile to the right, $j_{\phi 1}$ has a clear two-peak structure. The magnitude of the positive peak is still significantly larger than that of the negative peak because $j_c$ is still stronger than $j_I$. Such a two-peak structure of $j_{\phi 1}$ means that the integrated current $K$ as calculated by Eq. (25) has substantial cancellation between contributions from the positive and the negative peak, although the net value is still positive. The values of $K$ for different $\kappa$ with $q_0 = 0.7$ are shown in Fig. 5. It is substantially reduced and is less than the magnitude of $I$, the integrated current using $j_I$.

Next we consider a case with $q_0$ much closer to 1, the stability limit. Figure 8 shows a case with $q_0 = 0.9$, $\kappa = \kappa_\rho = 200$, and other parameters the same as in Fig. 6. From the left panel, it seems like there is a large difference in $\psi_1$ between the step-function case (dotted curve) and the continuous case (solid curve). Note however that the difference in the $r > a$ side is mainly a parallel shift, except within the boundary layer, such that the asymptotic $B_{\theta 1}$ (proportional to $\partial \psi_1 / \partial r$) of the two cases still agree well. In the right panel we see that $j_{\phi 1}$ has a very clear two-peak structure, with the negative peak stronger than the positive peak now. However, the total integrated current is still positive but small ($K \sim 0$), as can also be seen in Fig. 5 ($K \ll -I$). This is because while the positive peak is smaller in amplitude, it is wider than the negative peak and thus the contribution from both of them almost cancels each other. This is a clear example of why a step-function calculation can be misleading. Even though the surface current $K$ calculated for the step-function case is small, this does not necessarily mean that the first-order current density $j_{\phi 1}$ is small. In the present case, $j_{\phi 1}$ is not small by itself, only that it has internal structure of both signs within the boundary layer such that the integrated current becomes small. $j_I$ in this case is clearly shifted to the right of $j_c$. The two only have a small overlapping portion along $r$ such that the positive peak of $j_{\phi 1}$ is roughly given by $j_c$, while the negative peak is roughly given by $j_I$.

In cases with strong $j_I$ that is comparable with $j_c$ but shifted to the right, the background current density $j_{\phi 0}$ is no longer being transported by the flow as in Eq. (28). In fact, for positive $\xi_r$ so that $j_I$ is opposite to $j_{\phi 0}$, the leading edge of $j_{\phi 0}$ is being reduced by $j_I$. The end effect is that $j_{\phi 0}$ does not move as a whole in the direction of $\xi_r$. Instead, its profile just becomes steeper due to the fact that a part of it is moved forward by $j_c$, while the part outside the forward moving part is pushed back by $j_I$. However, the density profile is being transported by the flow velocity and moves forward in $\xi_r$ due to the continuity equation, assuming incompressibility. This means that the profiles of $j_{\phi 0}$ and $\rho$ will become different, with the profile of $\rho$ going beyond that of $j_{\phi 0}$ in the direction of positive $\xi_r$, even if they are the same initially. To take this effect into account properly, one would need a nonlinear (or quasi-linear) treatment, beyond the scope of the present calculation. Staying within the realm of the linear theory, we now consider a case with different profiles for $j_{\phi 0}$ and $\rho$ initially, with $\rho$ going beyond and covering $j_{\phi 0}$. Figure 9 shows the shifted density case with $q_0$ = 0.7, $a$ = 0.5, $a_\rho$ = 0.55, $\kappa = \kappa_\rho$ = 200, and other parameters remaining the same as in Fig. 6. Due to the fact that $j_c$ is proportional to $\partial j_{\phi 0}/\partial r$, while $j_I$ is proportional to $\partial \rho/\partial r$, as indicated in Eqs. (30) and (31), $j_c$ (dashed curve, overlapping the positive peak of $j_{\phi 1}$, the solid curve) and $j_I$ (dotted cure, overlapping the negative peak of $j_{\phi 1}$) are clearly separated. Both the magnitude of the peak and the width of $j_c$ are larger than that of $j_I$ so that the net current is positive, although there is still substantial cancellation between the two. The clear separation between $j_c$ and $j_I$ implies that the inner current density $j_{\phi 0}$ is totally transported by the flow velocity, according to Eq. (28). One subtlety is that now $j_I$ is large at the location where $j_{\phi 0}$ is very small. This means that the linear treatment is questionable even at a very small perturbation amplitude. Again, as discussed above, a nonlinear (or at least quasi-linear) treatment is needed in principle for cases with $j_{\phi 0}$ and $\rho$ of different profiles. We should point out that the reason we choose $q_0 = 0.7$ for the shifted density case is because we have not found unstable modes for $q_0 = 0.8$, and 0.9 using these profiles. We can further confirm this by considering the step-function limit with shifted density. As it turns out, analytical treatment as in Sec. II is possible only for the case with $m = n = 1$, due to the fact that the solution as in Eq. (32) only exists in this case. With this choice, the growth rate $\gamma$ is given by

$$\frac{\gamma^2}{2B_0^2} = \frac{\left(\frac{a^2}{a_\rho^2} - q_0\right)}{R^2 q_0^2} \left\{ 1 - \frac{\left(\frac{a^2}{a_\rho^2} - q_0\right)\left(1 + \frac{2}{\gamma \tau_w}\right)}{1 - \frac{a_\rho^2}{b^2} + \frac{2}{\gamma \tau_w}} \right\}. \tag{33}$$

Note that Eq. (33) tends to Eq. (14) for $m = n = 1$, in the special case of $a = a_\rho$. From Eq. (33), we see that the stability limit is shifted to $q_0 < (a/a_\rho)^2 < 1$. Note that this stability limit is the same as requiring $k_\parallel < 0$ for $r < a_\rho$ by Eq. (7). In our special case with $a = 0.5$, and $a_\rho = 0.55$, the stability limit is thus $q_0 < 0.826$. This explains why we have not found an unstable mode for $q_0 = 0.9$. For $q_0 = 0.8$, we still cannot find an unstable mode because this is too close to the stability limit and thus it would require a very sharp density boundary (extremely large $\kappa_\rho$) to have a small enough $\rho$ at the point where $k_\parallel = 0$ ($r \sim 0.559$), as we discussed above regarding Fig. 2. We need to point out that such a stability boundary can also change with time since now the evolution of $j_{\phi 0}$ and $\rho$ profiles is not the same due to the separation of $j_c$ and $j_I$. Again, a proper treatment of such evolution would require nonlinear physics. Nevertheless, if a strong current density similar to $j_I$ can be induced at the density boundary under a fully nonlinear treatment, such a current density can have important implications for how the plasma current flows into the wall during a disruption.

VI DISCUSSION AND CONCLUSION

In this paper, we have generalized the step-function treatment of external kink instability to arbitrary background current and plasma density profiles. We have presented our calculations of kink eigenmodes using continuous profiles in order to resolve confusion in the literature on the nature and sign of surface current based on step-function profiles. We have shown that the integration of the first-order current density, $j_{\phi 1}$, of the unstable mode agrees with the surface current $K$ calculated in Ref. 9, in the step-function limit. However, we have also shown that the surface current found in Refs. 5 and 8, named $I$ in this paper, also has the physical meaning as the residual current, after subtracting the convective current, i.e., the part of the first-order current responsible for the convection of the background current density. We conclude that this confusion, especially on the sign of the surface current, is not due to a technical error in any of papers involved. Rather, it is mainly due to the identification of different physical quantities by the same name, i.e., surface current, by different researchers. A factor that has added to the confusion is the use of step-function profiles in previous analyses. As we have shown, such confusion is resolved by looking at results based on continuous profiles.

We should remark that while step-function background profiles are commonly used in analytical theories, due to simplicity in analysis, they are idealization of the real system. Current density and plasma density profiles in real experimental plasmas generally do not have sharp (step-function like) boundaries. Moreover, in many direct 3D dynamical simulations, such as M3D, it is difficult to treat sharp boundaries within the plasma due to numerical reasons. Even in theories using step-function profiles, important physics might be missing within boundary layers. In the examples considered in this paper, we have shown that even if the surface current $K$ is zero in the step-function calculation, it does not mean that the first-order current density $j_{\phi 1}$ is zero. It can simply have an internal structure with both signs that cancel its integrated value. Linear analysis based on step-function profiles also has a possible mathematical consistency problem if there are physical quantities involved that are based on the spatial derivative of these profiles, like the $j_c$ and $j_I$ in Eqs. (30) and (31). In the step-function limit these quantities become infinite and thus can make the linear assumption problematic. Physically it might mean that nonlinear effects can be important at a very early stage of the linear instability. In view of all these considerations, we conclude that our analysis and numerical method is useful in calculating eigenmodes using general background profiles that can be compared with more realistic experimental and/or dynamical simulation data, in addition to providing physical understanding that can be missing in the step-function analyses.

Finally, we discuss possible physical implications of our results, especially on the physics of disruption. We have confirmed that the residual current density, $j_I$, can be strong and separate from the main current density in some cases, such as near the stability limit (Fig. 8) and the case in which the plasma density profile extends beyond the background current density profile (Fig. 9), similar to what have been proposed in literature.[5-7] The fact that $j_c$ and $j_I$ can be separated spatially also contributes to the possibility that the background current and plasma density can evolve into different profiles even if they are the same initially. Therefore, for a case in which the plasma density extends well beyond the background current density and has its edge near the wall of the vacuum vessel, a sudden induction of strong $j_I$ on the density edge during a disruption can have an important damaging effect if $j_I$ can interact with the wall. However, we caution that this implication is based on our linear analysis, which might be invalidated due to the fact that strong $j_I$ appears at locations where the background current density is small. Therefore, it is important to study this situation more carefully using direct nonlinear numerical simulations.[16] This is left to future work. What is evident from our work is that halo (or Hiro) currents need not be characterized by one sign or the other during the evolution of disruptions, and continuous plasma profiles of the type realized in numerical simulations or experiments can have a considerable spatial structure even within boundary layers that need to be taken into account in quantitative calculations of the forces on the wall.

## ACKNOWLEDGMENTS

This work was supported by a DOE Contract No. DEAC02-76CH03073. C.S.N. acknowledges the sabbatical support from the Princeton Plasma Physics Laboratory and the University of Alaska Fairbanks, and the support from a National Science Foundation Grant No. PHY-1004357.

Figure Captions

FIG. 1. Normalized profiles used in calculations with $\kappa$ = 20, 40, 60, 80, 100, and 200. The curve with the sharpest gradient corresponding to $\kappa$ = 200, while the gradient for $\kappa$ = 20 is much milder. The right panel plots over a smaller range of $r$ to show the boundary region better.

FIG. 2. $k_\parallel$ as functions of $r$ using current density profiles of Eq. (24), with $m = n = 1$, $R = 3$, $\kappa$ = 40, $q_0$ = 0.1, 0.2, 0.25 (dashed curve), 0.3, 0.4, 0.5, 0.6, 0.7, 0.8, and 0.9 (in the order from bottom to top). The dotted line shows $k_\parallel = 0$, which is also the $k_\parallel$ curve for $r < a$ in limit of $q_0$ tending to 1.

FIG. 3. Symbols show $q_0$ and $\kappa$ for all cases for which unstable eigenmodes are found. The solid curve indicates that $\rho$ at the point $k_\parallel = 0$ is less than $0.5\,\mathrm{erfc}(5) \sim 7.7\mathrm{e}{-13}$ for $\kappa$ above it.

FIG. 4. Symbols show growth rate $\gamma$ for different $q_0$ and $\kappa$, with $m = n = 1$, $R = 3$, and $\tau_w$ = 1000. The solid (dotted) curve in the left (right) panel shows $\gamma$ for the step-function case.

FIG. 5. The solid curve shows $K$ given by Eq. (20) with $\xi_r = -1$. The data points on it are calculated from Eq. (25) for all non-step-function cases, with $\xi_r = -1$ at $r = a$. The dashed curve shows $-I$ by Eq. (21). The data points on it are given by subtracting data of $K$ from $j_{\phi 0}\xi_r$.

FIG. 6. Eigenmode solution for $q_0 = 0.3$, $a = a_\rho = 0.5$, $b = 1$, $\kappa = \kappa_\rho = 40$, $m = n = 1$, and $R = 3$. The solid curve in the left panel shows $\psi_1$, while dotted curve shows the solution for the step-function case. The solid curve in the right panel shows $j_{\phi 1}$, while the dashed curve shows $j_c$ by Eq. (30), and the dotted curve shows $j_I$ by Eq. (31).

FIG. 7. Similar to Fig. 6 but for the case with $q_0 = 0.7$, and $\kappa = \kappa_\rho = 100$.

FIG. 8. Similar to Fig. 6 but for the case with $q_0 = 0.9$, and $\kappa = \kappa_\rho = 200$.

FIG. 9. Similar to Fig. 6 but for the case with $q_0 = 0.7$, $a = 0.5$, $a_\rho = 0.55$, and $\kappa = \kappa_\rho = 200$.

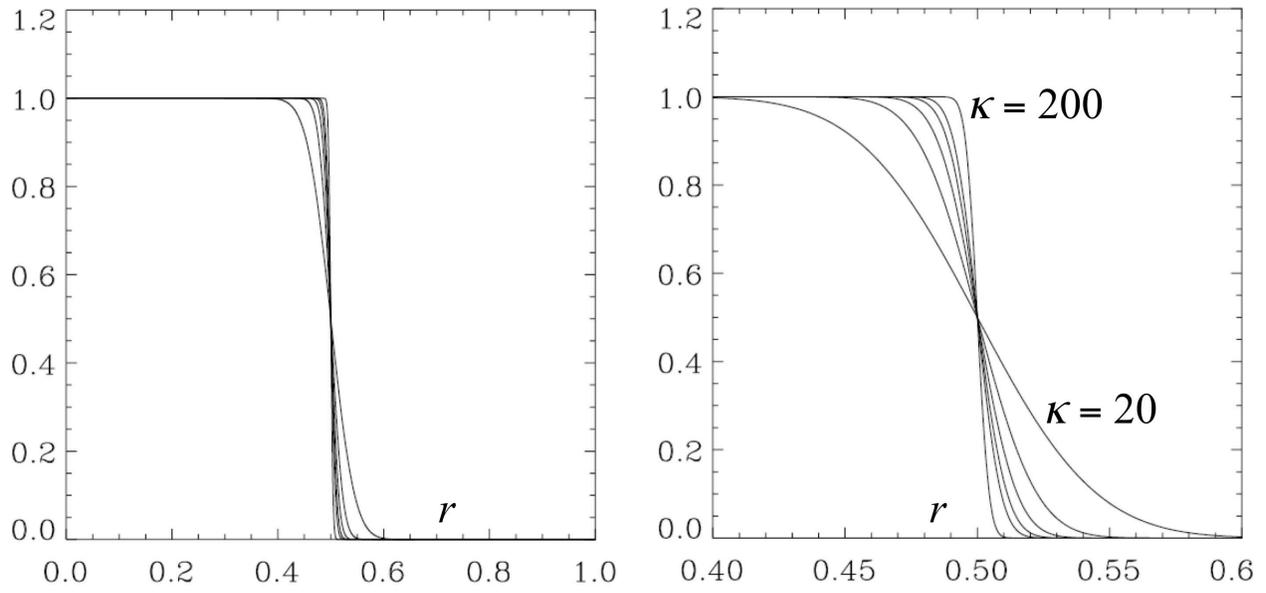

FIG. 1. Normalized profiles used in calculations with $\kappa$ = 20, 40, 60, 80, 100, and 200. The curve with the sharpest gradient corresponding to $\kappa$ = 200, while the gradient for $\kappa$ = 20 is much milder. The right panel plots over a smaller range of $r$ to show the boundary region better.

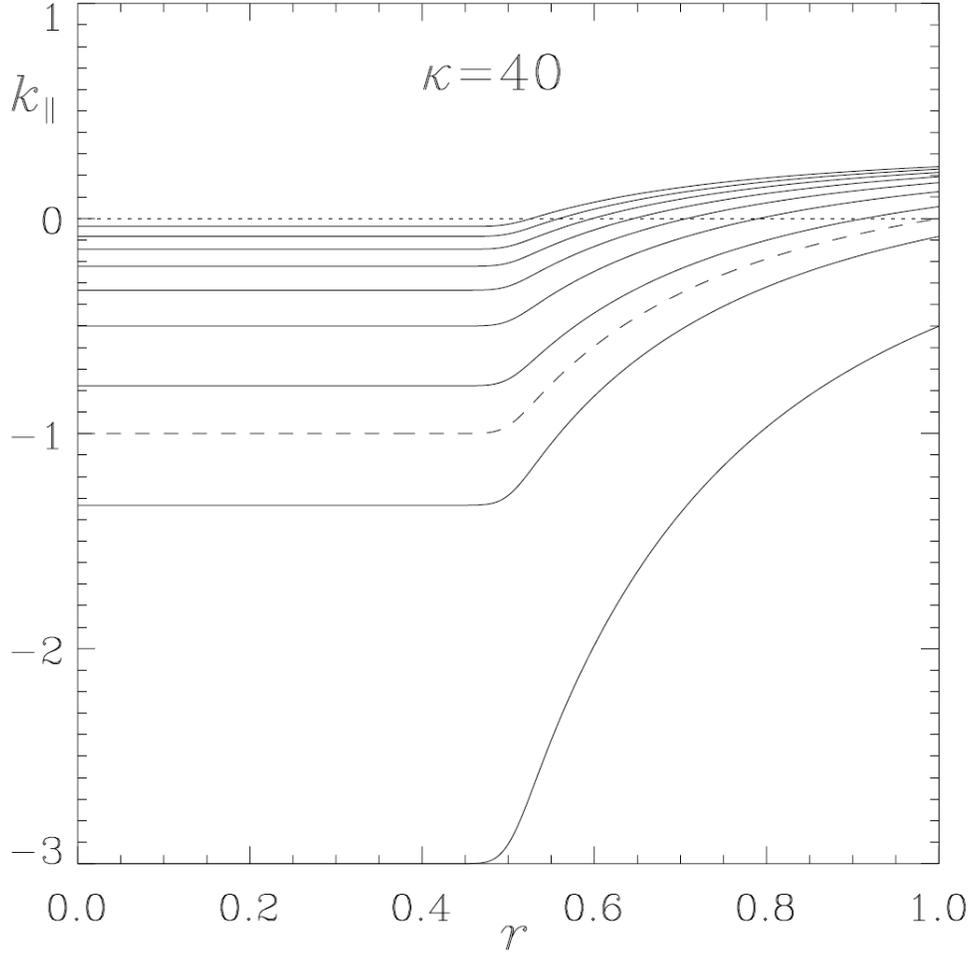

FIG. 2. $k_\parallel$ as functions of $r$ using current density profiles of Eq. (24), with $m = n = 1$, $R = 3$, $\kappa = 40$, $q_0 = 0.1, 0.2, 0.25$ (dashed curve), $0.3, 0.4, 0.5, 0.6, 0.7, 0.8$, and $0.9$ (in the order from bottom to top). The dotted line shows $k_\parallel = 0$, which is also the $k_\parallel$ curve for $r < a$ in limit of $q_0$ tending to 1.

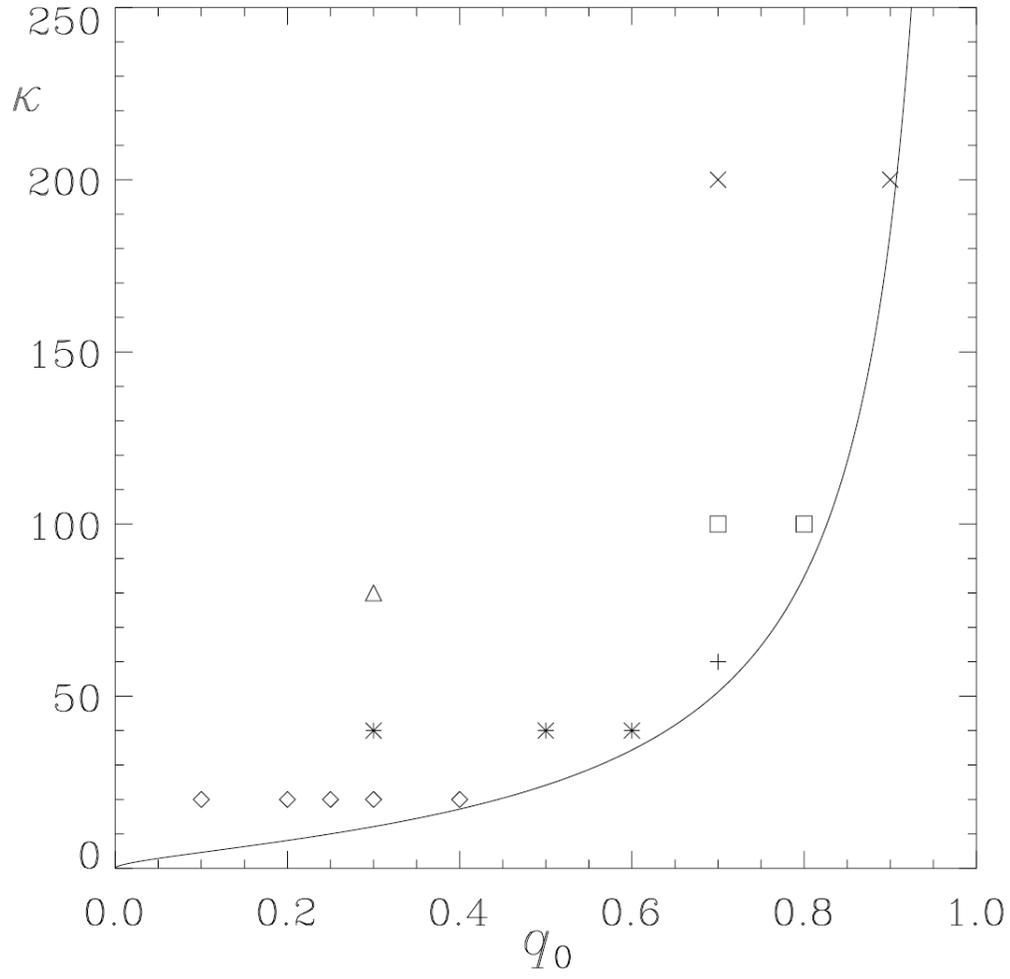

FIG. 3. Symbols show $q_0$ and $\kappa$ for all cases for which unstable eigenmodes are found. The solid curve indicates that $\rho$ at the point $k_\parallel = 0$ is less than 0.5erfc(5) ~ 7.7e-13 for $\kappa$ above it.

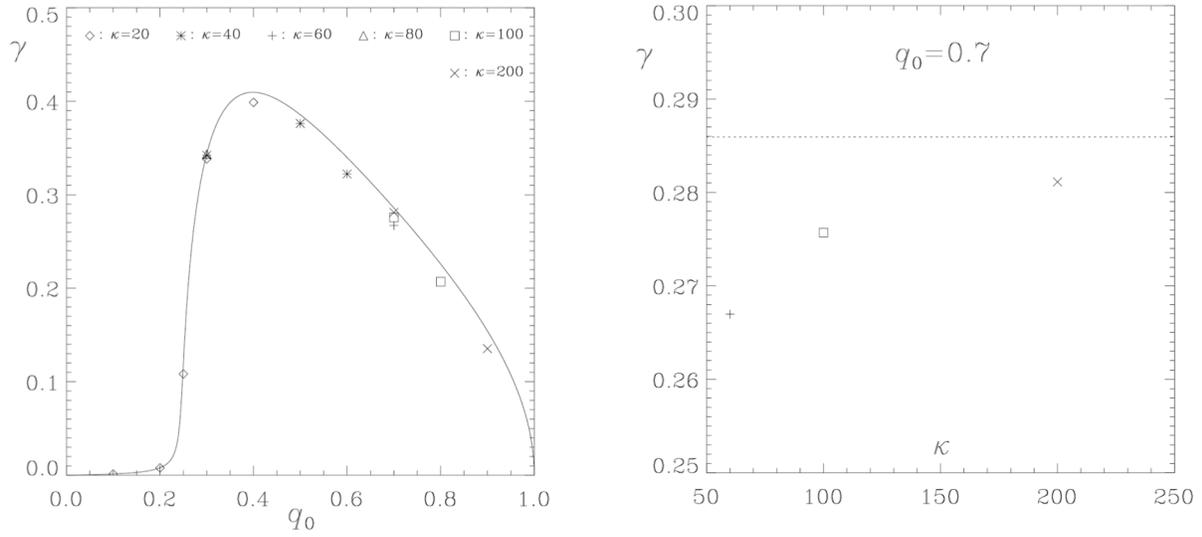

FIG. 4. Symbols show growth rate $\gamma$ for different $q_0$ and $\kappa$, with $m = n = 1$, $R = 3$, and $\tau_w = 1000$. The solid (dotted) curve in the left (right) panel shows $\gamma$ for the step-function case.

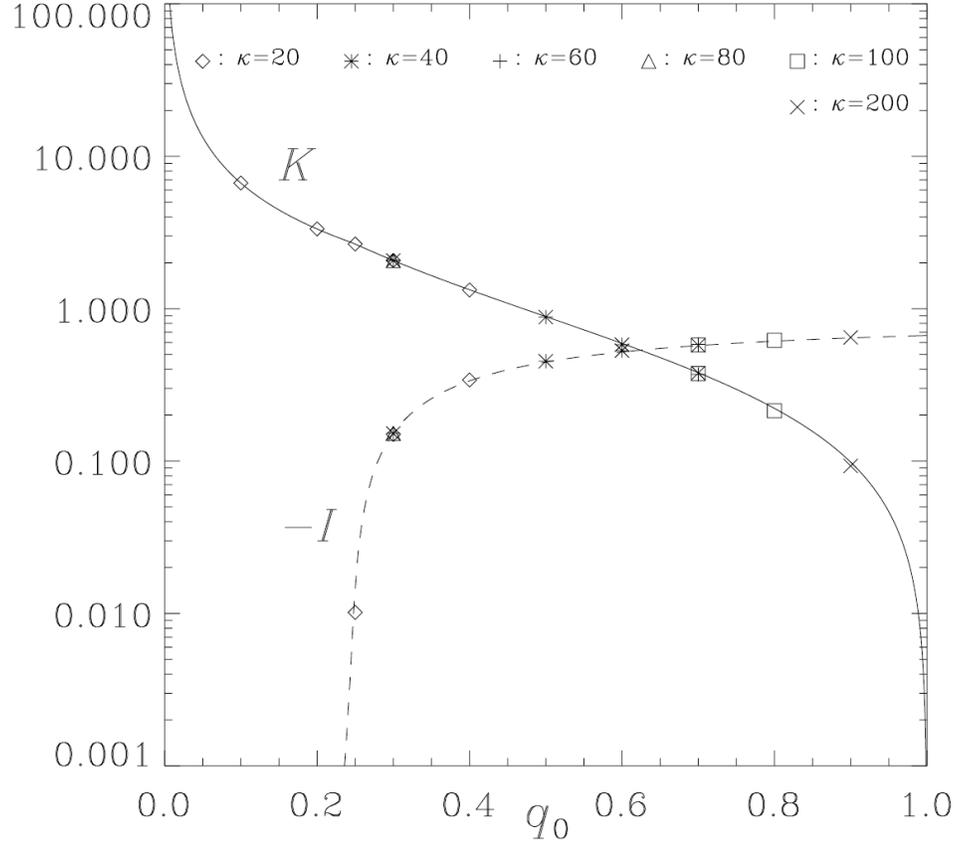

FIG. 5. The solid curve shows $K$ given by Eq. (20) with $\xi_r = -1$. The data points on it are calculated from Eq. (25) for all non-step-function cases, with $\xi_r = -1$ at $r = a$. The dashed curve shows $-I$ by Eq. (21). The data points on it are given by subtracting data of $K$ from $j_{\phi 0}\xi_r$.

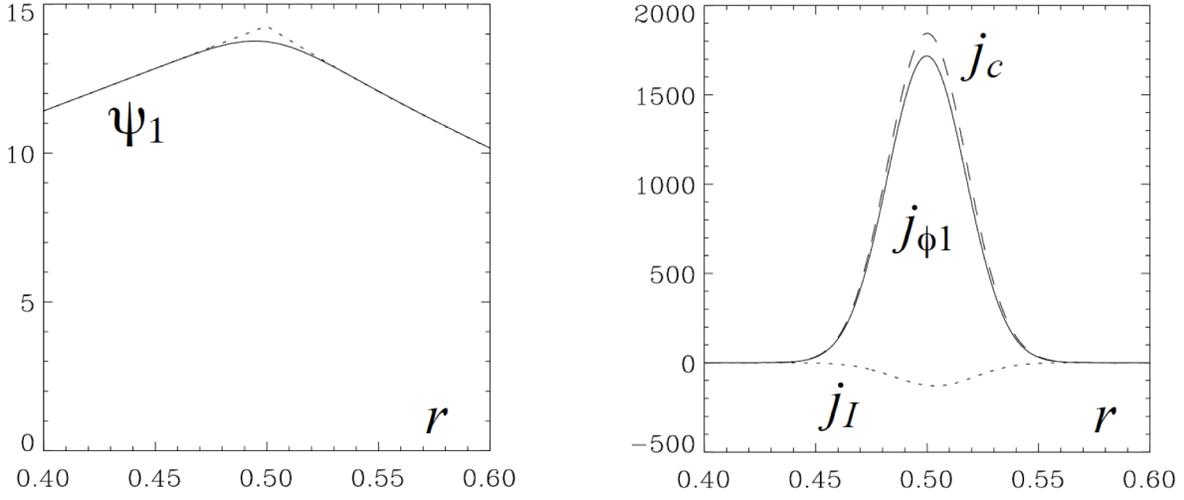

FIG. 6. Eigenmode solution for $q_0 = 0.3$, $a = a_\rho = 0.5$, $b = 1$, $\kappa = \kappa_\rho = 40$, $m = n = 1$, and $R = 3$. The solid curve in the left panel shows $\psi_1$, while dotted curve shows the solution for the step-function case. The solid curve in the right panel shows $j_{\phi 1}$, while the dashed curve shows $j_c$ by Eq. (30), and the dotted curve shows $j_I$ by Eq. (31).

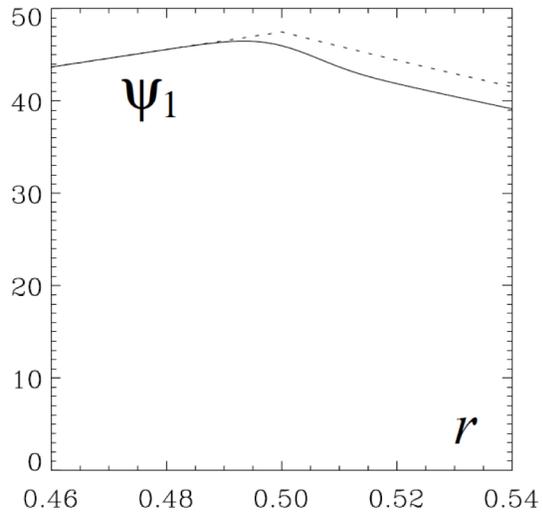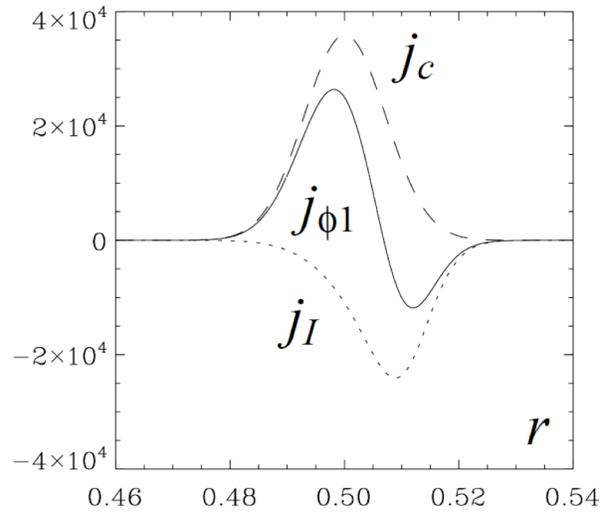

FIG. 7. Similar to Fig. 6 but for the case with $q_0 = 0.7$, and $\kappa = \kappa_\rho = 100$.

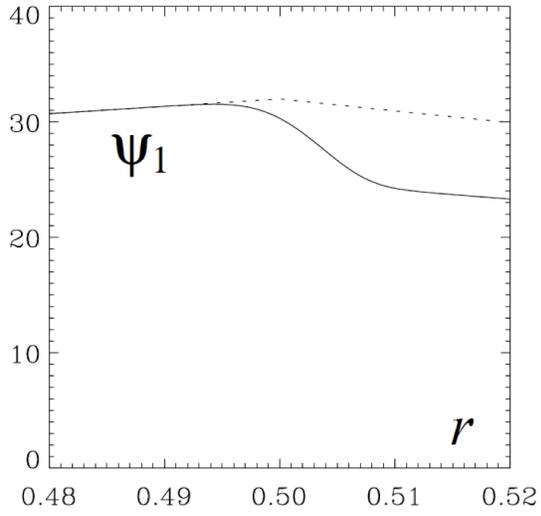 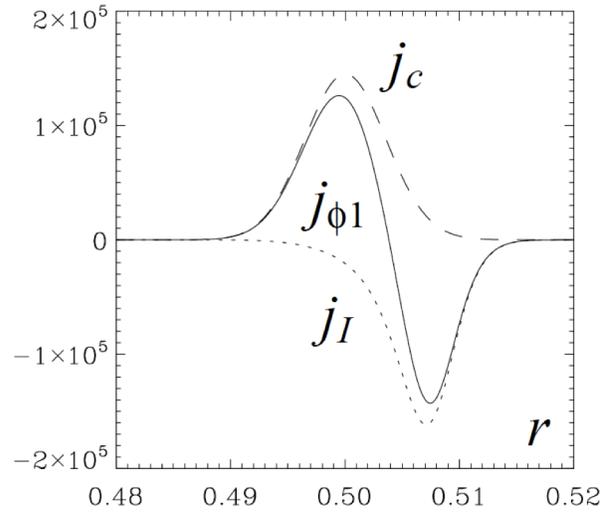

FIG. 8. Similar to Fig. 6 but for the case with $q_0 = 0.9$, and $\kappa = \kappa_\rho = 200$.

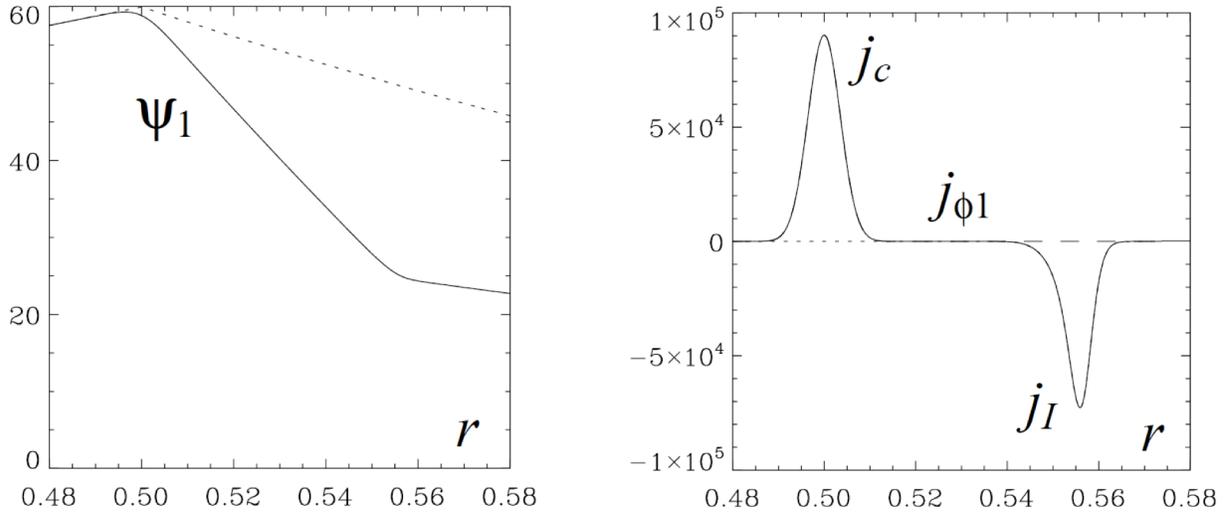

FIG. 9. Similar to Fig. 6 but for the case with $q_0 = 0.7$, $a = 0.5$, $a_\rho = 0.55$, and $\kappa = \kappa_\rho = 200$.